\documentclass{article}

\usepackage{amsfonts}
\usepackage{amsmath}
\usepackage{amsthm}
\usepackage{geometry}
\newtheorem{theorem}{Theorem}
\DeclareMathOperator*{\argmin}{argmin}

\title{A Note on Spectral Clustering and SVD of Graph Data}
\author{
Ziwei Zhang \\
Tsinghua University \\
zw-zhang16@mails.tsinghua.edu.cn \\
}

\begin{document}

\maketitle

\begin{abstract}
Spectral clustering and Singular Value Decomposition (SVD) are both widely used technique
for analyzing graph data. In this note, I will present their connections using simple linear
algebra, aiming to provide some in-depth understanding for future research.
\end{abstract}

\noindent \textbf{Notations} In this paper, a graph is $G=(V,E)$ where $V$ is a set of $N$ nodes and $E \subseteq V \times V$ is a set of edges. I only consider undirected graphs, so the adjacency matrix $\mathbf{A} \in \mathbb{R}^{N \times N}$ is symmetric. The graph is also assumed to be connected. The Laplacian matrix is defined as $\mathbf{L} = \mathbf{D} - \mathbf{A}$, where $\mathbf{D}$ is a diagonal degree matrix $\mathbf{D}_{i,i} =
\sum_{j \neq i} \mathbf{A}_{i,j}$. One normalized Laplacian matrix is defined as
$\mathbf{L}_{rw} = \mathbf{D}^{-1} \mathbf{L} = \mathbf{I} - \mathbf{D}^{-1} \mathbf{A}$, which is related to random walks on the graph. From linear algebra \cite{chung1997spectral}, both $\mathbf{L}$ and $\mathbf{L}_{rw}$ have non-negative real eigenvalues and real eigenvectors, and the eigenvalues of $\mathbf{L}_{rw}$ lie in [0,2]. The Singular Value Decomposition (SVD) of any matrix $\mathbf{R}$ is defined as $\mathbf{R} = \mathbf{U} \mathbf{S} \mathbf{V}^T$, where $\mathbf{S}$ is a diagonal matrix of singular values sorted in descending order $\mathbf{S}_{1,1} \geq \mathbf{S}_{2,2} \geq ... \geq \mathbf{S}_{N,N}$ and $\mathbf{U},\mathbf{V}$ are corresponding singular vectors. The eigenvalue decomposition (EVD) is defined as: $\mathbf{R} = \mathbf{X} \mathbf{\Lambda} \mathbf{X}^T$ , where $\mathbf{\Lambda}$ is a diagonal matrix of eigenvalues sorted in descending order according to the absolute value $ \left| \mathbf{\Lambda}_{1,1} \right| \geq \left| \mathbf{\Lambda}_{2,2} \right| \geq ... \geq \left| \mathbf{\Lambda}_{N,N} \right|$ and $\mathbf{X}$ are the corresponding eigenvectors.\\

\noindent \textbf{Spectral Clustering}  In this paper, I consider the spectral clustering proposed in \cite{shi2000normalized}, which aims to minimize the Normalized Cut. The spectral clustering algorithm is as follows \cite{von2007tutorial}: first compute $\mathbf{X}_{sc} \in \mathbb{R}^{N \times k}$, the $k$ eigenvectors of $\mathbf{L}_{rw}$ corresponding to the $k$ smallest eigenvalues
$\mathbf{\Lambda}_{sc}$, i.e. the last $k$ columns of $\mathbf{X}$ and $\mathbf{\Lambda}$; then, perform k-means on $\mathbf{X}_{sc}$ to get the clustering results.\\

\noindent \textbf{Singular Value Decomposition (SVD)} From Eckart-Young theorem \cite{eckart1936approximation}, the top-k SVD corresponds to the optimal rank-k decomposition of a matrix in terms of the Frobenuis norm. Formally,
\begin{equation}
    \mathbf{U}_k \mathbf{S}_k \mathbf{V}_k^T = \mathbf{P}^* = \argmin_{\substack{\mathbf{P} \\ s.t. \; rank(\mathbf{P}) = k}} \left\| \mathbf{P} - \mathbf{R} \right\|_F
\end{equation}
where $\mathbf{U}_k,\mathbf{S}_k,\mathbf{V}_k$ are the first $k$ columns of $\mathbf{U},\mathbf{S},\mathbf{V}$ respectively.\\

\noindent \textbf{Connection between SVD and EVD} From linear algebra, the following theorem can connect
SVD and EVD:
\begin{theorem}\label{thm1}
If a matrix has all real distinct eigenvalues and real eigenvectors, then
\begin{equation}
    \mathbf{U}_i\mathbf{S}_i\mathbf{V}_i^T = \mathbf{X}_i \mathbf{\Lambda}_i \mathbf{X}_i^T, 1\leq i \leq N.
\end{equation}
\end{theorem}
\noindent The theorem shows that the top-k EVD are basically the same as the top-k SVD (actually, except a few differences in the signs, i.e. putting the minus sign of eigenvalues into the singular vectors
since singular values are all non-negative). However, note that the top-k EVD are sorted according
to the absolute value, i.e. the largest ``magnitude'' instead of numerically largest. \\

\noindent \textbf{Connection between spectral clustering and SVD} Using the definition of $\mathbf{L}_{rw} = \mathbf{I} - \mathbf{D}^{-1} \mathbf{A}$, it is easy to obtain the following fact: spectral clustering features $\mathbf{X}_{sc}$ correspond to the eigenvectors of the largest eigenvalues of $\mathbf{A}_{rw} = \mathbf{D}^{-1} \mathbf{A}$. Then, it may be attempting to draw the conclusion that, $\mathbf{X}_{sc} \mathbf{\Lambda}_{sc} \mathbf{X}_{sc}^T$ is the best rank-k decomposition of $\mathbf{A}_{rw}$\footnote{A previous work \cite{tian2014learning} is based partially on this conclusion.}. However, such claim is \textbf{incorrect} or \textbf{incomplete} since one need to seek the top-k EVD according to the \textbf{absolute value} instead of numerically
largest values. In fact, the following theorem holds:
\begin{theorem}
$\mathbf{X}_{sc} \mathbf{\Lambda}_{sc} \mathbf{X}_{sc}^T$ is the best rank-k decomposition of $\mathbf{A}_{rw}$ if and only if top-k eigenvalues of $\mathbf{A}_{rw}$ are all positive.
\end{theorem}
\noindent However, this condition usually cannot hold true for real graphs. In fact, for real graphs, about half of the eigenvalues are positive while the other half are negative, as shown in \cite{zhang2018arbitrary}.

\noindent To seek a more general connection, graph signal processing technique is resorted to. As shown in \cite{shuman2013emerging}, EVD of the Laplacian matrix represents the ``smoothness'' of graph signals: the eigenvectors with small eigenvalues resemble smooth signal bases while eigenvectors with large eigenvalues represent signal bases that oscillate. Then, it is easy to obtain the following result using the fact that eigenvalues of $\mathbf{L}_{rw}$ are in range [0,2]: the smallest eigenvalues of $\mathbf{L}_{rw}$ correspond to the largest positive eigenvalues of $\mathbf{A}_{rw}$ (approximately near 1), while the largest eigenvalues of $\mathbf{L}_{rw}$ correspond to the smallest negative eigenvalues of $\mathbf{A}_{rw}$ (approximately near -1). Combining these two parts turns out to be the top-k eigenvalues of $\mathbf{A}_{rw}$, i.e. largest in absolute value. Using Theorem \ref{thm1}, they are exactly the SVD of $\mathbf{A}_{rw}$, which leads to the following conclusions:
\begin{itemize}
\item Spectral clustering corresponds to the most smooth signal bases of the graph.
\item SVD of $\mathbf{A}_{rw}$ corresponds to both the most smooth and most non-smooth signal bases of the graph.
\end{itemize}

\noindent \textbf{Conclusion} In this note, I analyze the connection between spectral clustering and SVD, two popular approaches applied to graph data, and the conclusion is highlighted above. This result
may be interesting to the research of Graph Convolutional Networks (GCNs) since spectral method
in filtering graph signals is widely adopted \cite{bruna2014spectral}.

\section*{Acknowledgment}
I thank Fei Tian at Microsoft Research Asia for helpful discussions.
\bibliographystyle{unsrt}
\bibliography{cite}

\end{document}